\documentclass[preprint,11pt]{article}
\pdfoutput=1

\usepackage{jcappub}
\usepackage{amsmath,latexsym,amssymb,color,epsfig}
\usepackage{dcolumn}
\usepackage{bm}
\usepackage{caption}
\usepackage{array}
\newcommand{\eq}[1]{(\ref{#1})}
\newcommand{\be}{\begin{equation}}
\newcommand{\ee}{\end{equation}}
\newcommand{\bea}{\begin{eqnarray}}
\newcommand{\ena}{\end{eqnarray}}
\newcommand{\no}{\noindent}
\newcommand{\nb}{\nonumber}

\renewcommand\a{\alpha}

\renewcommand\L{\ensuremath{\Lambda}}
\newcommand\m{\ensuremath{\mu}}

\newcommand\n{\ensuremath{\nu}}

\newcommand{\de}{\partial}

\newcommand{\ba}{\begin{eqnarray}}
\newcommand{\ea}{\end{eqnarray}}
\newcommand{\plm}{M_{\text{Pl}}} 
 
\title{Self-gravitating $ \pmb \Lambda$-media}

\author[a]{Marco Celoria}
\author[b]{Denis Comelli}
\author[c,1]{Luigi Pilo}

\affiliation[a]{Gran Sasso Science Institute (INFN)\\Via Francesco
  Crispi 7, L'Aquila, I-67100}
\affiliation[b]{INFN, Sezione di Ferrara,  I-35131 Ferrara, Italy}
\affiliation[c]{Dipartimento di Scienze Fisiche e Chimiche, Universit\`a di L'Aquila,  I-67010 L'Aquila, Italy}
\affiliation[1]{INFN, Laboratori Nazionali del Gran Sasso, I-67010 Assergi, Italy}
\emailAdd{marco.celoria@gssi.infn.it}
\emailAdd{comelli@fe.infn.it}
\emailAdd{luigi.pilo@aquila.infn.it}

\abstract{
 We address the question whether a medium featuring $p+\rho =0$, dubbed 
$\Lambda$-medium, has to be necessarily a cosmological constant.
By using effective field theory, we show that this is not the case for  a class of
media comprising perfect fluids,  solids  and special super solids, providing an
explicit construction.  The low energy
excitations are non trivial and lensing, the growth of large scale
structures can be used  to clearly distinguish $\Lambda$-media from  a
cosmological constant.}

\begin{document}
\maketitle
\flushbottom

\section{Introduction}
We do not know yet the nature of dark energy, though we have some
information on its equation of state. Actually, assuming a constant equation of state, observations indicate $p+\rho \approx 0$, which
points toward the simplest possibility: a cosmological constant (CC). 
Suppose 
that the {  upcoming} large scale structure surveys will
establish that $w\equiv p/\rho\equiv -1$,  can we conclude then that dark energy behaves as a CC? 
For a pure CC we have   $\rho=-p =$constant and the corresponding fluctuations are zero:
$\delta\rho=\delta p=0$; moreover gravitational waves propagate with a
massless dispersion relation $\omega^2=k^2$. The goal of this paper is
to show that is possible to construct  {  simple field
theory models based on an action principle}, describing what we call a  $\Lambda$-medium featuring
$p+\rho = 0$ as a non-perturbative equation of
state. Though in a Friedman-Lemaitre-Robertson-Walker (FLRW)
background  a $\Lambda$-medium is completely equivalent to a CC this
is not the case when perturbations are introduced. Indeed, we
still have   $\delta
p=-\delta \rho$,  but with non-trivial
perturbations. 
\\
The starting point is  the assumption that 
dark energy can be effectively described as an isotropic medium whose low-energy
excitations   are   phonon-like. Such
behaviour is rather common in condensed matter systems but also in
cosmology.

The outline of the paper is the following. In section \ref{SGM} we
briefly recall the basics of self gravitating media that represent the
general framework of which $\Lambda$-media are a very special
subset. $\Lambda$-media are introduced in section \ref{LM} together
with their thermodynamical properties. Section \ref{LMDS} is devoted
to the study of the conditions under which $\Lambda$-media are stable
under linear perturbations around flat space. Scalar cosmological
perturbation are studied in section \ref{LMC} while the tensor ones
are analysed in section \ref{LMGWs}. The study of how the growth of
structure is modified in the presence of $\Lambda$-media is given in
\ref{LMGS}. Finally, section \ref{CONC} contains our conclusions.

\section{Self-Gravitating Media}
\label{SGM}
The key tool we  use is the effective field theory (EFT) description of the
dynamics of media~\cite{Leutwyler:1993gf,Leutwyler:1996er, Son:2002zn, Son:2005ak, Dubovsky:2005xd,
  Dubovsky:2011sj,Ballesteros:2012kv,ussgf,classus} according to which 
 the low energy excitations (phonons) can be described by a classical
theory of four derivatively coupled scalar fields $\varphi^A$,
$A=0,1,2,3$. 
Among the various action principle formulations of the dynamics of
media~\cite{Taub:1954zz, Seliger1968, Schutz:1970my, Carter1973,
  Carter:1987qr, khalatnikov1982relativistic}, the EFT framework is
formulated in terms of unconstrained fields and it is
capable to describe perfect fluids, super fluids,  solids and
super solids depending on the internal symmetries of the action which
translates in the form of the energy momentum tensor (EMT).
A similar formalism was used in the contest of
inflation~\cite{Endlich:2012pz,Cannone:2014uqa,Graef:2015ova,Lin:2015cqa,Bartolo:2015qvr}, though barring any 
special relation among operators realising $p+\rho=0$.
Following the notations of~\cite{ussgf,classus},  the leading
operators in the EFT {for homogeneous media, invariant under the shift symmetry $\varphi^A\rightarrow \varphi^A+\lambda^A$ for constant $\lambda^A$,}  can be written in terms of the matrix
\be
C^{AB}  = g^{\mu \nu} \de_\mu  \varphi^A \de_\nu  \varphi^B \, , 
\ee
and the velocity fields $u_\mu$ and $\mathcal{V}_\mu$
\be
\label{velocities}
u^\mu = -\frac{\epsilon^{\mu \nu \alpha \beta}}{6 \, b\sqrt{-g} }\epsilon_{abc}
\, \partial_\nu \varphi ^a \, \, \partial_\alpha \varphi ^b \, \partial_\beta \varphi ^c \,
\qquad
\mathcal{V}_\mu=-\frac{\partial_\mu \varphi^0}{\sqrt{-X}}\, ,
\ee
where $g_{\mu \nu}$  is the space-time metric, $b=\sqrt{\det \left(B^{ij}
  \right)}$,  with $ B^{ij} \equiv C^{ij}$ and  $X=C^{00}$.  
  Small latin indices like $i,j$, assume the values 1,2,3 while
  greek and capital latin ones the values 0,1,2,3; we shall
  also  denote by $\pmb{B}$ the $3\times3$ matrix
with matrix elements $B^{ij}$.
Being $u^\mu \de_\mu \varphi^j =0$, $\varphi^j$ can be
interpreted as the spatial Lagrangian (comoving) coordinates of the medium,
while $\varphi^0$ represents the clock's medium.
The action of  a self-gravitating medium in the presence of  gravity is 
 \be
\label{fact}
 S=\plm^2\;\int d^4 x\;\sqrt{-g}\, R+\int d^4x\sqrt{-g} \; U\left(C^{AB}\right)\,;
 \ee
where $R$ is the Ricci scalar, $\plm=(16\pi G)^{-1/2}$ is the Planck mass and $U$ is
the medium Lagrangian depending on the derivative of the St\"uckelberg
fields $\varphi^A$.
At leading order we have a total of 9 operators $Y,\;X,\;\tau_{1,2,3},\;y_{0,1,2,3}$  (compatible with { isotropy, 
implemented as a}   global internal spatial $SO(3)$ symmetry {  $\varphi^i\rightarrow R^{i}_{\phantom{i}j} \varphi^j$, with the constant matrix $\mathbf{R}\in SO(3)$}) defined as 
\bea
Y= u^\mu \de_\mu \varphi^0 ,\qquad \tau_n = \text{Tr} \left(
  \pmb{B}^n \right),\qquad y_n=\text{Tr}\left(\pmb{B}^n \, \pmb{Z} \right)  
\ea
 where 
$\pmb{Z}$ is $3\times3$ matrix with matrix elements $Z^{ij}=C^{0i} C^{0j}$ 
(note that $b= (\tau_1^3-3\;\tau_2\;\tau_1+2\;\tau_3)/6$). Moreover, we define the  operators $w_n= \text{Tr} (\pmb{W}^n)$, where $\pmb{W}\equiv \pmb{B}-\pmb{Z}/X$.\\
Interestingly, the EFT formalism allows also to give a thermodynamical   
interpretation~\cite{usthermo, classus} by which  
some combinations of   operators  are related to  thermodynamical  variables. 

\section{$\Lambda$-Media}
\label{LM}
From the action \eqref{fact},   the  energy momentum tensor (EMT)  for
the most generic media  has the following structure
\begin{equation}
T_{\mu\nu}=-\frac{2}{\sqrt{-g}}\frac{\delta  S}{\delta g^{\mu\nu}}=
p\, h_{\mu\nu}+\rho\, u_\m u_\n+2\, q_{(\m}u_{\n)}+\pi_{\m\n}
\end{equation}
where
$\rho= u^\mu u^\nu T_{\mu\nu}$, $p=h^{\mu\nu}T_{\mu\nu}/3$,   $q_\mu$ satisfies $u_\mu q^\mu=0$ and $\pi^\mu_{\phantom{\mu}\mu}=u^\mu \pi_{\mu\nu}=0$. 
\\
Depending on the internal symmetries, we can select some special
combinations of the operators appearing in the Lagrangian,
corresponding to particular classes of media. For
  instance, fluids and  super fluids $U(X,Y,b)$ are protected by
  invariance under volume preserving diffeomorphisms 
\begin{equation}\label{vol_diff}
\varphi^a \rightarrow \psi^a(\varphi^b)\,,  \qquad  \text{det} \left|\partial \psi^a/\partial \varphi^b\right| = 1 \, .
\end{equation}
Since for $\Lambda$-media  $w=-1$, the condition
$p(U)=-\rho(U)$ at the non-perturbative level is equivalent to a differential equation for $U$, which
can be solved in terms of the basic operators considered as
independent variables.
\\
Let us describe this procedure   for the  following  classes of media 
\begin{itemize}
\item  solids, characterised by $U(Y,\tau_n)$
\item special super solids, characterised by $U(X,w_n)$
\item perfect fluids, characterised by $U(Y,b)$.
\end{itemize}  
 Solids are selected by the invariance under  the  internal symmetry~\cite{Dubovsky:2004sg,ussgf}
\begin{equation}\label{symm1}
\varphi^0 \to \varphi^0 + f(\varphi^j)
\end{equation}
and the thermodynamical  dictionary is given in Table \ref{table:Dictionary}.
The entropy per particle $\sigma=s/n $ is constant in time
\begin{equation}
\dot{\sigma}=u^\mu\partial_\mu \sigma=0 \, .
\end{equation}
From the non-perturbative expression for $p$ and $\rho$ in  Table
\ref{table:Dictionary}, imposing   $w = -1$ gives
\begin{equation}
 Y\;U_Y-\frac{2}{3}\sum_{m=1}^3m\;\tau_mU_{\tau_m}=0\, ,
\end{equation} 
which reduces the original   dependence of $U$ from the
  original four operators down to the
following  three  special combinations 
\begin{equation}
U^{\Lambda}_{S}=U\left(Y\tau_1^{3/2}, \frac{\tau_2}{\tau_1^2},\frac{\tau_3}{\tau_1^3} \right)
\end{equation}
  invariant  under the Lifshitz scaling
  \cite{Dubovsky:2004ud,Endlich:2012pz, Celoria:2016vul}
\begin{equation}\label{Lifshitz}
\left\{
  \begin{array}{lr}
   \varphi^0 \rightarrow  &\lambda^{-3}\, \varphi^0\\
    \varphi^j \rightarrow  & \lambda \, \varphi^j
  \end{array}
\right. \, .
\end{equation}
An interesting subcase is the isentropic   solid with an entropy per
particle constant in spacetime, described by
$U(\tau_n)$ (where only the spatial St\"uckelberg $\varphi^j$ are
present) and     characterised by $s=0$. In particular,
$\Lambda$-isentropic solids are described by 
\begin{equation}
U^{\Lambda}_{ IS}=U\left(\frac{\tau_2}{\tau_1^2},\frac{\tau_3}{\tau_1^3 }\right)\, .
\end{equation}
Special super solids $U(X,w_n)$  are selected by   the
  invariance under   the  internal symmetry
  \cite{Dubovsky:2004sg,Rubakov:2008nh,ussgf,classus}
\begin{equation}\label{symm2}
\varphi^j\rightarrow \varphi^j +f^j(\varphi^0) \, .
\end{equation}
$\Lambda$-special super solids are obtained   imposing $p=-\rho$ for
the expressions in Table \ref{table:Dictionary};
we have 
\begin{equation}
U^{\Lambda}_{ SS}=U\left(Xw_1^{3},
  \frac{w_2}{w_1^2},\frac{w_3}{w_1^3}\right) \, ,
\end{equation}  
which is again invariant under the Lifshitz scaling \eqref{Lifshitz}.
Similar media were studied in a cosmological setting  in
\cite{Dubovsky:2004ud,Dubovsky:2005dw} and also in the contest of spherically
symmetric solutions in massive gravity~\cite{Comelli:2010bj,Bebronne:2009mz}.
\\
Isentropic special super solids,
described by $U(w_n)$, are  invariant under \eqref{symm2}
and~\cite{ussgf}
\begin{equation}\label{symm3}
\varphi^0\rightarrow \varphi^0 +f(\varphi^0) \, .
\end{equation}
Remarkably, a proposal for a UV completion for isentropic special super solids involving at LO the scalar operators $w_1$ and $w_2$ was put forward in \cite{Blas:2014ira}, where the  temporal St\"uckelberg $\varphi^0$ is embedded into the khronometric model and the spatial St\"uckelberg $\varphi^j$ are coupled  to a triplet of Higgs vector fields.
\\
The symmetry \eqref{symm3} forbids the operator $X$ in $U^{\Lambda}_{SS}$ and we conclude that  $\Lambda$-isentropic special super solids are described by 
\begin{equation}
U^{\Lambda}_{ISS}=U\left(\frac{w_2}{w_1^2},\frac{w_3}{w_1^3}\right)\, .
\end{equation} 
Finally,   perfect fluids are protected by the symmetries \eqref{vol_diff} and
 \eqref{symm1};  their Lagrangian is of the form  $U(Y,b)$.
 \\
$\Lambda$-perfect fluids with $p+\rho=0$ have the following Lagrangian~\cite{ussgf} 
\begin{equation}
U^{\Lambda}_{PF}=U(Yb)
\end{equation} 
which is protected by the enhanced symmetry
\begin{equation}
\varphi^A \rightarrow \psi^A(\varphi^B)\,,  \qquad  \text{det} \left|\partial \psi^A/\partial \varphi^B\right| = 1 \, .
\end{equation}
The $\Lambda$-perfect fluid EMT simply reads 
$T_{\mu\nu}=p \,g_{\mu\nu}$  whose conservation leads directly to $\rho=$constant. 
Although  a   $\Lambda$-perfect fluid is similar to a CC,
the non-vanishing entropy per particle  indicates  that underlying degrees of freedom are present.
Moreover, the St\"uckelberg fields satisfy non-trivial equations of
motion in order to keep  the combination $b \, Y$ constant. 
\\
Differently from fluids, anisotropic stress in solids allows to have a conserved energy-momentum tensor,
$p+\rho=0$   and non-trivial gradient for the pressure and
  the energy density.
 Actually, for a solid
$
T_{\mu\nu}=p\;g_{\m\nu}+\pi_{\m\n}
\;$
 where  $\pi_{\m\n}$ is  the anisotropic  stress; from the
 conservation  of the energy-momentum tensor  we get
\bea\label{dSSolid}
p=-\rho,\qquad \dot p = \pi_{\m\n}\;\nabla^\n\,u^\m=\pi_{\m\n}\;\sigma^{\m\n},
\qquad 
{\cal D}_\m p=-{\cal D}^\n\pi_{\m\n}-a^\n\pi_{\m\n} 
\ea
where $u^\a\nabla_\a p\equiv \dot p$  and  we have used the
  standard 
  decomposition of the covariant derivative of the velocity in terms of
   rotation and shear according with
$\nabla_\n\,u_\m=\sigma_{\n\m}+\omega_{\n\m}+ \nabla_\alpha
u^\alpha/3\;h_{\n\m}-u_\n\;a_\m$ and ${\cal D}^\m=h^\m_\n\,\nabla^\n$,
$a^\mu = u^\nu \nabla_\nu u^\mu$.
Remarkably, the relations  \eqref{dSSolid} are intrinsically different from the
 corresponding relations for a CC
\be
\label{CCeq}
 p=-\rho=-\Lambda,\qquad\nabla_\mu\,p=0\;\;\Rightarrow\;\; p=-\Lambda \, .
 \ee 
For a FRW background   (\ref{dSSolid}) and (\ref{CCeq}) coincide,
while  for a perturbed  FRW  metric deviations from a CC are present  already
  at the first order. 

\begin{table}
\centering
\begin{tabular}{||c|c|c|c|| }
\hline 
 & $U(Y,\tau_n)$&  $U(X,w_n)$& $U(Y,b)$ \\
\hline
$\rho$ & $ -U+Y\;U_Y$& $-U+2XU_X$&$ -U+Y\;U_Y$  \\
\hline
$p$ & $U-\frac{2}{3}\sum_{m}m\;\tau_mU_{\tau_m}$& $U-\frac{2}{3}\sum_{m}m\;w_mU_{w_m}$& $U-b\;U_b$  \\
\hline 
$n$ & $\tau_m^{3/(2m)}$ & $w_m^{3/(2m)} $  & $b$ \\
\hline 
$s$ & $ U_Y$ &$-2\sqrt{-X}U_X$ &$ U_Y$   \\ 
\hline 
\end{tabular} 
\caption{Thermodynamical correspondence}
\label{table:Dictionary}
\end{table}
\section{Dynamical Stability}
\label{LMDS}
Since $\Lambda$-media are very particular, it is important to
  study  their stability. Though self-gravitating media can
  have a familiar  Jeans-like instability at some scale $k_J$,  no
instabilities should be present at very large $k$.  
In this regime, curvature and the mixing of the
St\"uckelberg fields with gravity are negligible and, much like in  spontaneously
broken gauge theories, the ultraviolet  behaviour is captured by
the St\"uckelberg fluctuations. One can  forget about gravity and simply study the effective
quadratic Lagrangian obtained expanding $U$ around
  Minkowski space
\be\label{pert_stuck}
\varphi^0=t+\pi_0(t,\vec{x}),\ \ \varphi^j=x^j+V^j(t,\vec{x})+\partial^j
\pi_L(t,\vec{x}) 
\ee
with $\partial_j V^j=0$. As discussed in~\cite{classus}, the
dynamics of linear perturbations is controlled by five 
parameters $\{M_b\}$ (with $b=0,1,..,4$), expressed in terms of first and
second partial derivatives of the Lagrangian $U$ with respect to the
basic operators~\footnote{In this paper we use the same definition for  $\{M_b \}$
as in~\cite{Celoria:2017hfd} which differs from the one in~\cite{classus} by a
factor $a^4$; namely $M_b$ in~\cite{classus} is equal to $a^4 \, M_b$ in the present paper.}, their
expressions are given in the appendix. These parameters are related to  the  mass terms of the
metric fluctuations $h_{00}$, $h_{0i} $ and $h_{ij}$  in the unitary
gauge appearing in  rotational invariant massive gravity, see \cite{Rubakov:2004eb,Rubakov:2008nh,Comelli:2013txa} and \cite{Comelli:2014xga} for the non-perturbative structure. 
\\
The propagation of scalar perturbations is controlled by  two mass parameters: if $\hat{M}_0\neq 0 $ and $\hat{M}_1+ \bar{p}+ \bar{\rho}\neq 0$  the total energy in the scalar sector is given by
\be
\begin{split}
E_s=  \hat{M}_0 \; \pi _0'{}^2+\frac{k^2}{2} \, 
   \left(\hat{M}_1+ \bar{p}+ \bar{\rho}  \right)\; \pi _L'{}^2+ k^4 \, 
   \left(\hat{M}_2-\hat{M}_3\right)\; \pi _L^2- \frac{k^2}{2}
   \hat{M}_1 \,  \pi _0^2 \, .
\end{split}
\ee
When the kinetic term of  $\pi _0$ or  $\pi _L$ vanishes, at least one degree of freedom can be integrated out and the stability analysis has to be redone. For details see \cite{Celoria:2017hfd}.
\\
Notice that no condition on $p$ and $\rho$ has been imposed. 
The Lagrangian for transversal vectors $\pmb{V}$ perturbations  reads
\be
L_V =\frac{1}{2} \left(\hat{M}_1 + p+ \rho  \right) \pmb{V}'^2  
-\hat{M}_2 \, k^2 \, \pmb{V}^2 \, .
\ee
Imposing that energy is  bounded from below in both the scalar and
vector sectors leads to
\be
\begin{split}
&M_0  \geq  0  \, , \qquad   \mathcal{M}_1 \equiv \hat{M}_1  + \bar{p}+ \bar{\rho} \geq0 \, , \\
&M_1 \leq 0  \, , \qquad M_2  \geq M_3   \, , \qquad M_2 \geq0.
\end{split}
\label{stabc}
\ee
In the limit  $p+\rho=0$, combining \eqref{massrel} with
\eqref{stabc} we get
\begin{equation}\label{cond}
 M_{2} >M_3\geq \frac{3}{2} \, M_0\geq 0 \,  ,\qquad M_1=0 \, .
\end{equation}
The very same stability conditions are obtained by expanding the action (\ref{fact}) around a
generic FLRW space-time in the Newtonian gauge at the quadratic order and
imposing stability in the limit $k \to \infty$ \cite{Celoria:2017hfd}. 
Remarkably,  the condition  $M_1 =0$ is protected by symmetries
\cite{Dubovsky:2005xd,ussgf} and thus stable $\Lambda$-media are
adiabatic \eqref{entropy_prop}.
Gradient and  ghost  instabilities are absent  also for media
characterised by $M_0 =0$ or $M_2 =M_3 $
\cite{Dubovsky:2004sg, Dubovsky:2005xd}.
While solids tend to be sensitive to the
introduction of higher operators, this is not the case
for special supersolids~\cite{Dubovsky:2004sg,Rubakov:2008nh}.
A detailed analysis of the sixth mode in
massive gravity and self-gravitating media is given in \cite{Celoria:2017hfd}.

\section{ Cosmology of  $\Lambda$-media}
\label{LMC}
The  fluctuations of  $\Lambda$-media around  de Sitter (dS)
space-time are particularly interesting and  show many connections
with modified gravity theories.
\\
 In the Newtonian gauge, using conformal time, the scalar
 perturbations of the metric are
\be
ds^2=a(t)^2 \, \left[(-1 + 2 \,    \Psi )\, dt^2+(1+2 \, 
\Phi )\, d\vec{x}^2   \right]\, ;
\ee
while for $\varphi^0$ we have
\be\label{pert_stuck2}
\varphi^0=\phi(t)+\pi_0(t,\vec{x}),
\ee
and $\varphi^j$ is expanded as in \eqref{pert_stuck}.\\
For a generic medium, at  the background level  entropy per particle
is conserved and pressure and energy density enter
in the standard Friedmann equations.
\\
For  $\Lambda$-media, we have
\begin{equation}\label{deltarho0}
\delta p=-\delta \rho=
   -\frac{\phi'\,M_4}{a^4\,M_0}\;\delta\sigma \, ;
\end{equation}
where 
\begin{equation}\label{entropy_pert}
  \delta \sigma = 2\, \frac{\hat{M}_0  }{\phi '}\;\left[
  \Psi+\frac{\pi_0'}{\phi'}-\frac{M_4}{M_0}\;(3\; \Phi+k^2\;\pi_L)
  \right] \,
\end{equation}
and $\hat{M}_i=\plm^2\;M_i$.
The Friedmann equations and
$\mathcal{H}'' = 2 \, \mathcal{H}^3$, valid when $\bar{p}=-\bar{\rho}$, together with
\eqref{deltarho0}, give the following relations among the mass parameters 
\be
M_0=M_4,\quad
M_2=3\;(M_3-M_4) \, .
\label{massrel}
\ee
Notice that  $\{M_b\}$ are constant parameters for $\Lambda-$media.
The very same relations also follow  from the invariance under \eqref{Lifshitz}. 
For $\Lambda$-media the function $\phi$  appearing in \eqref{pert_stuck2} is determined by the
conservation of the EMT at the background level and the relations
among the mass parameters \eqref{massrel}. Thus, $\Lambda$-media
select naturally a de Sitter (dS) background for which 
\be\label{bkg}
a(t) = - \frac{1}{H_0  \, t} \, ,\quad {\cal
  H}=\frac{a'}{a}=-\frac{1}{t} \, ;
\ee
where $H_0 $ is an integration constant. Notice that at the background level the  conservation of the
$\Lambda-$medium EMT is equivalent to 
\be
\phi'= \phi_0 \; a^4 \, ;
\label{phidd}
\ee
where $\phi_0$ is { an  integration} constant.
From  \eqref{deltarho0} and \eqref{phidd}, we have
\begin{equation}
\label{deltarho}
\delta p=-\delta\rho=-\phi_0\; \delta\sigma \, ,
\end{equation}
where, as consequence of the conservation of  $\Lambda-$medium
EMT. The constant $\phi_0$ simply rescale $\delta \sigma$ and can be
set to 1.  the  entropy  per particle perturbations satisfy 
\begin{equation}\label{entropy_prop}
\delta \sigma'= \frac{a^4 \, \hat{M}_1}{  \phi'^2}\;\;k^2\;(\pi_0-\phi'\;\pi_L') \, .
\end{equation}
When $\delta\sigma'=0$, i.e. the entropy per particle is a function of
spatial coordinates only  $\delta\sigma(\vec x)$, the
medium is  {\it adiabatic}; this is the case when $M_1=0$, see  \eqref{entropy_prop}.
 When the stronger condition $\delta\sigma=0$ is imposed, the medium
 is  {\it isentropic} and  $M_{0,4}=0$, see \eq{entropy_pert}. Notice
 that perturbations are essentially entropic as a result of $\delta p
 = - \delta \rho$.
 
From the stability conditions we have seen that  stable
$\Lambda$-media are adiabatic or isentropic;  
in the following we will focus on the cases $M_1=0$ and
$M_{0,1}=0$. At the linear order, the EMT of an adiabatic, namely
$M_1=0$,  $\Lambda$-medium has the following form
\bea
&&T^{(1)}_{00} = a^2 \left(\phi_0 \, \delta \sigma  - 2 \, \bar \rho \, \Psi
\right)\, ; \nb \\
&&T^{(1)}_{0i} =0 \, ;\\
&&T_{ij}^{(1)}= 2  \, \hat M_2 \, a^{2} \, \de_i \de_j \pi_L
-\frac{\delta_{ij}}{2} a^2 \left(3 \, \delta \sigma  \, \phi _0+
4 \, \Phi  \, \bar \rho\right) \, .
\nb
\ena
The  conservation
of the $\Lambda-$medium EMT   only gives  that $\delta \sigma$ is
constant in time, see (\ref{entropy_prop}),  but also that it is related to $\pi_L$ by
\be
3 \, \, \phi _0 \, 
  \delta \sigma  + 4 \, \hat M_2 \, k^2 \, \pi _L =0 \, ;
\label{consm}
\ee
where  we have switched to Fourier space setting $k^2 = k^i k^j
\delta_{ij}$, with $k^i$ is the comoving momentum. 
The scalar sector of the linear perturbed Einstein equations for a generic $\Lambda$-medium  reads
\be
\begin{aligned}
\label{eqEE}
a^2\, \delta \sigma  &= 4\;\plm^2\;\left[ k^2\;\Phi+3\;{\cal H}\;(\Phi'+{\cal H}\;\Psi)\right]\, ,\\
-\frac{\phi'^3\; \delta \sigma'}{4\, a^2\, \plm^2}&=  k^2 \; \left(\Phi'+{\cal H}\;\Psi\right)\, ,
\\
a^2 \, M_2   \;\pi_L \,  &=   \left(\Psi- \Phi\right)\, ,
 \end{aligned}
\ee
where  $\delta\rho$ is given by \eqref{deltarho0} together with  \eqref{entropy_pert} and \eqref{bkg}, for $\delta\sigma'$ see \eqref{entropy_prop}.
{\it $\Lambda$-perfect fluids}  $U^{\L}_{ PF} $ for which $M_2=M_1=0$,  behave  exactly as a CC,
indeed 
\be
\delta\rho=\delta\sigma=\delta p= \Phi=\Psi=0,\quad
\pi_0'=k^2\;\phi'\;\pi_L \, .
\ee
What differentiate such a fluid from the CC is the presence non-trivial
perturbations of the St\"uckelberg fields and  a
constant entropy $\bar{\sigma}$; unless the $\Lambda$-perfect fluid is
directly coupled with matter \cite{Celoria:2017xos}, no physical effect is present.
For instance, during  $\Lambda$-perfect fluid domination,  a
subdominant dark matter sector has a constant  density  contrast
$\delta_{m}=\delta \rho_{m}/\bar \rho_{m}$ as for the  case of a CC.
\\
{\it $\Lambda$-solids} $U^\L_{S}$ and {\it $\Lambda$-special super solids}    $U^\L_{SS}$ are characterised by the same
 structure of perturbations  and $\delta\sigma=\delta \sigma_0(k)$ so
 that
\begin{equation}
\begin{aligned}\label{relationsBadeen}
 \delta\rho&=-\delta p=  \delta\sigma=
  \text{const.}\, ,\\
 \Phi&=\Phi_0\;a^2,\qquad \, \frac{\Psi}{2}+\Phi =0
\end{aligned}
\end{equation}
and $\Phi_0=\frac{\delta\sigma}{4\;k^2\,\plm^2}
$.
Remarkably, the expression of $\Phi$ is universal and  is determined
by the constant value of the entropy per particle perturbation
$\delta\sigma_0$.

One can also check that vector modes do not propagate and
tensor modes have a massive dispersion being  $M_2 \neq 0$.
\\
For {\it $\Lambda$-isentropic solids} $U^\L_{IS}$ and
  {\it$\Lambda$-isentropic special super solids}    $U^\L_{ISS}$ the
  entropy density vanishes, $s=0$ and the  behaviour  of the scalar perturbations on dS are  the same as in
the case of CC domination being $M_1=M_0=0$. 
The only  detectable difference is that  tensor modes have a massive
dispersion relation being $M_2 \neq 0$. 
The features of stable $\Lambda$-media with $M_1=0$ are summarised in table \ref{table:dof}.
\\
Finally, consider what happens when the stability conditions
\eqref{stabc} are not satisfied.  Take {\it
  $\Lambda$-super fluids} characterised by $M_1\neq0 $ and $M_0\neq
0$. From the invariance under \eqref{vol_diff}, the anisotropic stress
$\pi_{\mu\nu}$ is zero, and  thus $M_2=0$. From the Einstein equations \eqref{eqEE} the two Bardeen potentials are equal $\Phi=\Psi$ and  satisfy the simple equation
\ba
 \Phi''-k^2\;\Phi=0,\;
\ea
with a general solution
\be
\Phi=\Phi_1\;e^{\frac{k}{a H_0}}+\Phi_2\;e^{-\frac{k}{a H_0}} \, .
\ee
The entropy per particle perturbation is given by
\ba
\delta\sigma=\frac{4\plm^2}{a^2}\;\left[3\;{\cal H}\;\Phi'+\left(k^2+3\;{\cal H}^2\right)\;\Phi\right] \,.
\ea
Though, the time behaviour of $\Phi$ is under control, at large $k$
there is an exponential grow. 
 
\begin{table}
\centering
\begin{tabular}{||c|c|c|c|c|c||}
\hline 
$\Lambda$-Medium& $\delta\rho=-\delta p,\;\delta\sigma$   &$\Psi$&$\Phi$&$\delta_m$&$m_g$\\ [.1cm]  
\hline 
CC  & $ 0$ &  $ 0$&$  0$&$\text{const.}$&$0$\\[.1cm]  
\hline 
$U^\L_{PF}$   &$ 0$ &  $ 0 $&$  0$&$\text{const.}$&$0$\\ [.1cm]  
\hline 
 $U^\L_{S}\; \&\;U^\L_{SS}$   &$\delta \rho_0=\delta\sigma_0$ &  $-2\;\Phi$&$ \Phi_0\; a^2$&$\propto a^2$&$\neq0$\\ [.1cm]
\hline 
 $U^\L_{IS}\; \&\;U^\L_{ISS}$   &$ 0  $ &  $   0$&$ 0$&$\text{const.}$&$\neq 0$\\ [.1cm]  
\hline 
\end{tabular} 
\caption{Features of  the different $\L$-Media. The quantities
  $\Phi_0$ and  $\delta \sigma_0$ are time independent.}
\label{table:dof}
\end{table}

 \subsection{Gravitational Waves}
\label{LMGWs}
The quadratic Lagrangian for  tensor perturbations  in
Fourier space is \cite{Dubovsky:2004ud,Rubakov:2008nh,Blas:2009my,ussgf}
\begin{equation}\label{GW}
L_T=\frac{a^2 \, \plm^2}{2} \left[
\; \chi_{ij}'^2- \chi_{ij}^2 \;\left(k^2 +a^2 \, M_2\right)\right ] \, ,
\end{equation}
where $\chi_{ij}$ is the   transverse and traceless spin two part of the metric perturbations.
For  perfect fluids and super fluids, where $M_2=0$, the dynamics of spin 2
modes is standard. 
This is not the case for  solids and super solids
where $M_2  \neq 0$. 
\\
If the accelerated expansion of the universe is related to the presence of the graviton mass then
 the graviton mass  has to be the of order of $m_g=\sqrt{M_2}\sim 10^{-33} \ \text{eV}$.
On the other hand, massive gravitons represent also a cold dark matter
candidate when $m_g \ge10^{-27} \ \text{eV}$ \cite{Dubovsky:2004ud}.
However,   bounds from gravitational waves observations  as GW150914
and the time delay of $1.7$ seconds between GW170817 and the
electromagnetic counterpart GRB 170817A  
led to $m_g \lesssim  \times 10^{-22} \ \text{eV}$. 

Let us comment briefly on the familiar Higuchi bound \cite{Higuchi:1986py,Higuchi:1986wu}.
Such a bound on the Pauli-Fierz mass in dS spacetime is derived when the massive spin 2 action gives no contribution to the background~\cite{Deser:2001wx,Grisa:2009yy}.
By definition, the above considerations do not apply for
self-gravitating $\Lambda$-media, where the dS background is related
to the energy density of the $\Lambda$-medium through  the Friedmann
equations.
\\
Note that for self-gravitating Lorentz invariant $\Lambda$-media, Lorentz
invariance implies $p+\rho=0$  and the relations
$M_1=M_2,\;M_3=M_4,\;M_0=M_3-M_1$ that, from \eq{massrel} and \eq{cond}, 
 requires $M_0=M_4=M_3,\;M_1=M_2=0$.  Thus the only
$\L$-media compatible with the above conditions are the $\L$-perfect
fluids, thus the graviton is still massless. 

\section{Modified Growth of Structure}
\label{LMGS}
Let us start by considering the evolution of  dark matter (DM)
  and dark energy in the form of an adiabatic  $\Lambda$-medium,
  particularly we focus on the two limits of dark matter domination
  and dark energy domination,   which can be treated
    analytically. The general case is studied numerically.
Taking dark matter as a perfect
fluid with equation of state $p_m=0$, the only (indirect)
coupling with the $\Lambda$-medium is via gravity. Being $M_1=0$, the
only contribution to $0i$ component of the perturbed EMT comes from
the DM fluid and the corresponding Einstein equation gives an equation
for the scalar velocity $v_m$ of DM; namely
\be
 a^2 \;v_m \, \bar \rho _{m} -4\; \plm^2 \;\left(\Phi '+ 
   \mathcal{H}\;\Psi\right) =0 \, .
\ee
The $00$ component of the perturbed Einstein equation allows to
express $\delta \rho_{m}$ in terms of the gravitational scalar
perturbations and $\delta \sigma$ as
\be
\begin{split}
&a^2\;\delta \rho_{m}=4\;\plm^2 \; \Phi  \left(  k^2 +3\;{\cal H}^2\right)-  \left( 1+9\;\frac{{\cal H}^2}{k^2}\right) \delta \sigma 
+ 12 \, \plm^2 \, \mathcal{H} \; \Phi '\, .
\end{split}
\ee
The remaining perturbed equations can
be casted in a second oder equation for $\Phi$
\be
\Phi'' +3 \, \mathcal{H} \, \Phi' -3\;w\;{\cal H}^2 \; \Phi
+ \frac{3\;{\cal H}^2}{4\;k^2\;\plm^2} \;(3\;w-2)\; \delta \sigma =0    \, ;
\ee
where
\be
{\cal H}^2 = \frac{a^2(\bar \rho_\Lambda+ \bar \rho_{m})}{6 \, \plm^2}
\, , \qquad \qquad 
w = -\frac{a^2 \, \bar \rho_\Lambda}{6 \, {\cal H}^2 \, \plm^2} \, ;
\ee
and  $\bar \rho_{m}$ and $\bar \rho_\Lambda$ are
the background dark matter  and dark energy density respectively.
During an expansion  phase dominated by an adiabatic
$\Lambda$-medium, one can check that by combining the continuity and
Euler equations for a  subdominant dark matter component its density
contrast behaves as $\delta \rho_{m} \sim a^2$, in sharp contrast with
$\delta \rho_{m}=$constant found in the familiar case of CC domination.
For $ \bar \rho_m\gg \bar \rho_\L $ we get $H^2\simeq \frac{H_0^2}{a}$ and the
leading terms are 
exactly those in DM 
dominated universe
\ba
 &&\Phi= \bar \Phi  \, ;
\\
&& \delta_{m}= 
\frac{2 \;a \;k^2\;
   \bar \Phi}{3\;H_0^2} \, +\bar \delta  \, ;
\ea
where we have neglected sub-leading decreasing modes and $\bar \Phi$,
$\bar \delta$ are integration constants. 
For $ \bar \rho_{\Lambda}\gg \bar \rho_{m}$ we get $H^2\simeq a^2\; H_0^2 $
 with  neglecting again sub-leading decreasing modes
\ba
&&\Phi = 
 \frac{ a^2 }{4}\;\frac{\delta\sigma}{k^2\;\plm^2}  \\
&& \delta_{m} = \bar \delta
-\left[\frac{3\;a^2}{4}+\frac{k^2\; \log(a)}{4\;H_0^2}\right]
\frac{\delta\sigma}{k^2\;\plm^2}  \, .
\ea
Moreover
\be\label{lean}
\Phi-\Psi=\frac{3}{4}\; a^2 \, \frac{\delta\sigma}{k^2\;\plm^2} \, .
\ee
The  universal and simple relations  \eqref{relationsBadeen} and \eqref{lean}  which
relate the Bardeen potentials   give  a clear prediction for lensing. 

\no

In a more realistic Universe, taking into account also radiation,  one can solve 
numerically, in the linear regime, the   conservation equations for photons
and dark matter plus a combination of the  Einstein equations, namely
\be
\begin{split}
4 \;k^2\; v_{\gamma }  &= 3\; \left(4 \;\Phi'+\delta _{\gamma }'\right)\, ; 
\\
4\; v_{\gamma }'&=4\; \Psi-\delta _{\gamma } \, ;
\\
k^2 \;v_m &= 3\; \Phi'+\delta _m'\, ; 
 \\
  v_m'&=\Psi-\mathcal{H}\; v_m\, ;
  \\
\frac{a^2\,  (\delta \rho_\Lambda +\delta \rho_\gamma +\delta \rho_m)}{2\,  M_{Pl}^2}&=\, 2 \, k^2 \, \Phi+6 \, \mathcal{H}( \Phi'+  \mathcal{H}  \, \Psi ) \, ;
   \\
    \Phi-\Psi &=\frac{3\; a^2 \; \delta\rho_\Lambda}{4 \; k^2 \; \plm^2 }\, .
\end{split}
\ee
where $\delta_\gamma=\delta\rho_\gamma/\bar\rho_\gamma$ and $\delta_m=\delta\rho_m/\bar\rho_m$ and recall that $\delta\rho_\Lambda=\delta\sigma=\ constant$.
The  linear matter  power spectrum is given in terms of the 2-point correlation
function for the gauge
invariant matter contrast $\Delta_m = \delta_m - 3 \, {\cal H}  \, v_m$.
The total density contrast $\Delta$ is
given by the weighted sum over the various components 
\be
\Delta= \sum_i \,  \frac{\bar\rho_i}{\bar\rho} \, \Delta_i \, ;
\ee
and satisfies a second order equation, see for instance~\cite{Celoria:2017xos}. 
\\
More precisely, the  matter power spectrum $P(k,z)$ is defined by 
\begin{equation}
\left< \Delta_m(\pmb{k},z) \Delta_m(\pmb{k}',z)\right>=\frac{P(k,z)}{(2\pi^3)} \delta(\pmb{k}+\pmb{k}') \, ,
\end{equation}
where    $< > $ are ensemble averages, and  can be determined once the initial conditions,   adiabatic  or entropic, are specified. As usual, using the Poisson equation, we can write $\Delta_m(z)$ in terms of the gravitational potential, which can be related to the  initial  gravitational potential, $\Phi_{in}=\Phi(z_{in})$, where $z_{in}=10^8$ in the following, by means of the transfer function. 

Adiabatic perturbations are characterized by $\delta\sigma=0$, together with the usual relations $\delta_\gamma=2 \,\Phi_{in}$, $\delta_m=3 \,  \Phi_{in}/2$ and $v_m=v_\gamma=\Phi_{in}/(2\mathcal{H}_{in})$.
\\
Finally,  observations from the CMB and LSS suggest, in agreement with the simplest inflationary models, that $\Phi_{in}$ is  a random field  drawn from a nearly-Gaussian distribution with mean zero and variance  distribution specified by the primordial power spectrum 
\begin{equation}
\begin{aligned}
P_\Phi&= \left< \Phi_{in}({\pmb{k}}) \, \Phi_{in}({\pmb{k}'}) \right>=\frac{\mathcal{A}_\Phi}{k^3}\left( \frac{k}{H_0}\right)^{n_s-1}
\end{aligned}
\end{equation}
where   $\mathcal{A}_\Phi$ and $n_s$ are chosen as in the $\Lambda$CDM model.
\\
Conversely, entropic initial conditions $\delta\sigma(z_{in})=\delta\sigma(z=0, \pmb{k})$ can be specified in terms of the dimensionless quantity $\delta_{\Lambda}=\delta \rho_\L/\rho_\L$ 
associated to the primordial power spectrum
\begin{equation}
P_\Lambda= \left< \delta_{\Lambda}({\pmb{k}}) \,\delta_{\Lambda}({\pmb{k}'}) \right>= 
  \frac{\mathcal{A}_{\Lambda}}{k^3} \; \left(
  \frac{k}{H_0}\right)^{n_\Lambda-1} \, .
\end{equation}
Since the mechanism that generated these intrinsic entropy
perturbations is not known, we regard the amplitude
$\mathcal{A}_{\Lambda}$ and the spectral index $n_\Lambda$ as free
parameters.   When perturbations originate from thermal
  rather than quantum fluctuations large deviation from a scale
  invariant  primordial power spectrums is possible,  see for
  instance  \cite{Magueijo:2002pg, Magueijo:2007na, Cai:2009rd, Biswas:2013lna}.
\\
Note that we neglect the possible existence of relative entropic perturbations   $\propto 4\;\delta_m-3\;\delta_\gamma$.

\begin{figure}[h!!]
\begin{center}
\includegraphics[width=13cm,height=7cm]{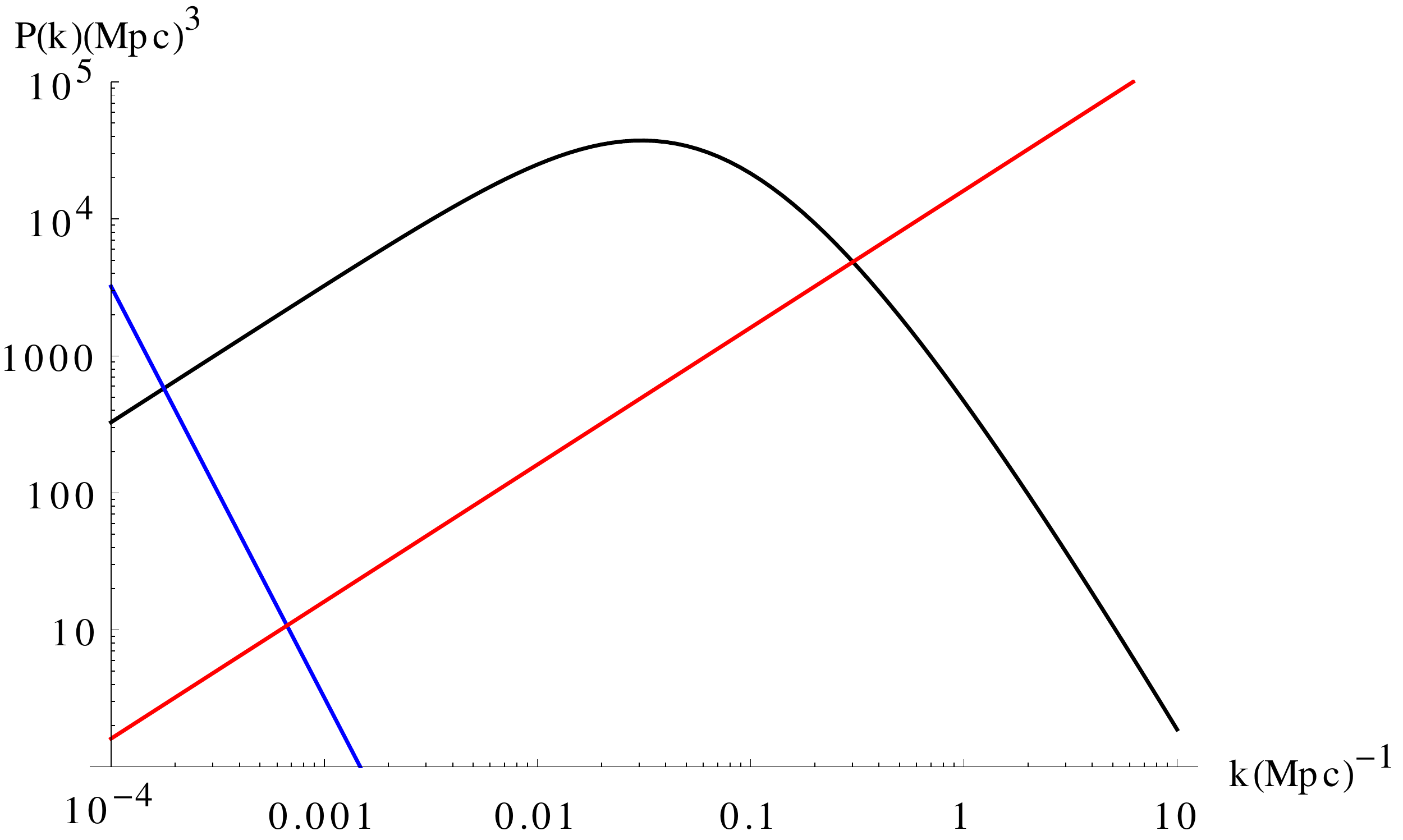}
\caption{Matter power spectrum for $\Lambda$CDM (black) ${\cal P}_{\L CDM}$ obtained by setting
  $\delta \sigma=0$. 
  The     blue   curve represents   ${\cal P}_\L$ with $n_\Lambda=1$ and the  red curve
  with $n_\Lambda=5$, with $\mathcal{A}_{\Phi}=\mathcal{A}_\Lambda$.
   We neglect
 the   cross correlation between adiabatic and entropic modes
 setting  ${\cal P}_{\Phi \Lambda}=0$.
 The total power spectrum { is the sum} of the black and red curve or the black and the blue one.}
\label{Figpower}
\end{center}
\end{figure}

Finally, the total matter power spectrum today (at $z=0$) can be expressed in terms of the primordial power spectrum schematically as
\begin{equation}
\begin{aligned}
P(k)&\sim  \left(\frac{k}{H_0}\right)^4 \left[ T_\Phi(k)^2  P_\Phi 
+T_\Lambda(k)^2  P_{\Lambda}
 +T_{\Lambda}(k)T_{\Phi}(k)  \left< \delta_{\Lambda}({\pmb{k}}) \, \Phi_{in}({\pmb{k}'}) \right>
\right] \\
 &\equiv {\cal P}_{\Lambda \text{CDM}}(k) + {\cal P}_{\Lambda}(k) + {\cal P}_{\Phi \Lambda}(k) \, 
 \end{aligned}
\end{equation}
where  $T_\Phi(k)$ and $T_\Lambda(k)$ represent the transfer functions for the two   possible initial conditions. 
\\
The choice of pure adiabatic initial conditions
corresponds to set $\delta_{\L}=0$; the matter spectrum is given just by
${\cal P}_{\Lambda \text{CDM}}$ and coincides with the 
$\Lambda$CDM power spectrum, the black dotted curve in Figure \ref{Figpower}. 
The new contributions  stem from ${\cal P}_{\Lambda}$ and $ {\cal P}_{\Phi \Lambda}$ and their impact depends on
  the primordial spectrum for $\delta \sigma$ and the correlation between adiabatic and entropic contributions.  
  \\
Specifically, while ${\cal P}_{\Lambda \text{CDM}}(k) \sim k^{n_s}$ for  $k\rightarrow 0$ and ${\cal P}_{\Lambda \text{CDM}}(k)\sim k^{n_s-4} $ on small scales (neglecting logarithmic corrections), we have $ {\cal P}_{\Lambda}(k)\sim k^{n_\Lambda-4}$ always.
Contrary to $\Lambda$CDM, where the change of slope is
due to modes that entered the horizon at matter domination or at
radiation domination, the time-independent nature of $\delta \sigma$
makes its contribution to the matter power spectrum basically
monotonic in $k$.
\\
The result is shown in Figure
\ref{Figpower} with $n_s=1$ for the adiabatic spectral index, while for the entropic index we consider the two cases $n_\Lambda=1$ and $n_\Lambda=5$ (matching ${\cal P}_{\Lambda \text{CDM}}$ at small or large scales, respectively), and we have chosen the same amplitude $\mathcal{A}_\Phi=\mathcal{A}_\Lambda$.
\\ 
 From Figure \ref{Figpower} it is clear that unless the size and the shapes of initial
perturbations for $\delta \sigma$ are tiny compared with the
primordial ones of $\Lambda$CDM, an excess of power at small or large
scale will appear, depending on the spectral index $n_\Lambda$.   

The presence of non-trivial perturbations in the dark sector 
changes the growth function $D(z)$  defined here as
\be
D(z)= \frac{ \Delta_m(z)}{ \Delta_m(z_{\text{late}})} \, , \qquad
z_{\text{late}}=10 \, .
\ee


\begin{figure}[h!]
\begin{minipage}{.5\textwidth}
\centering
\includegraphics[width=.8\linewidth]{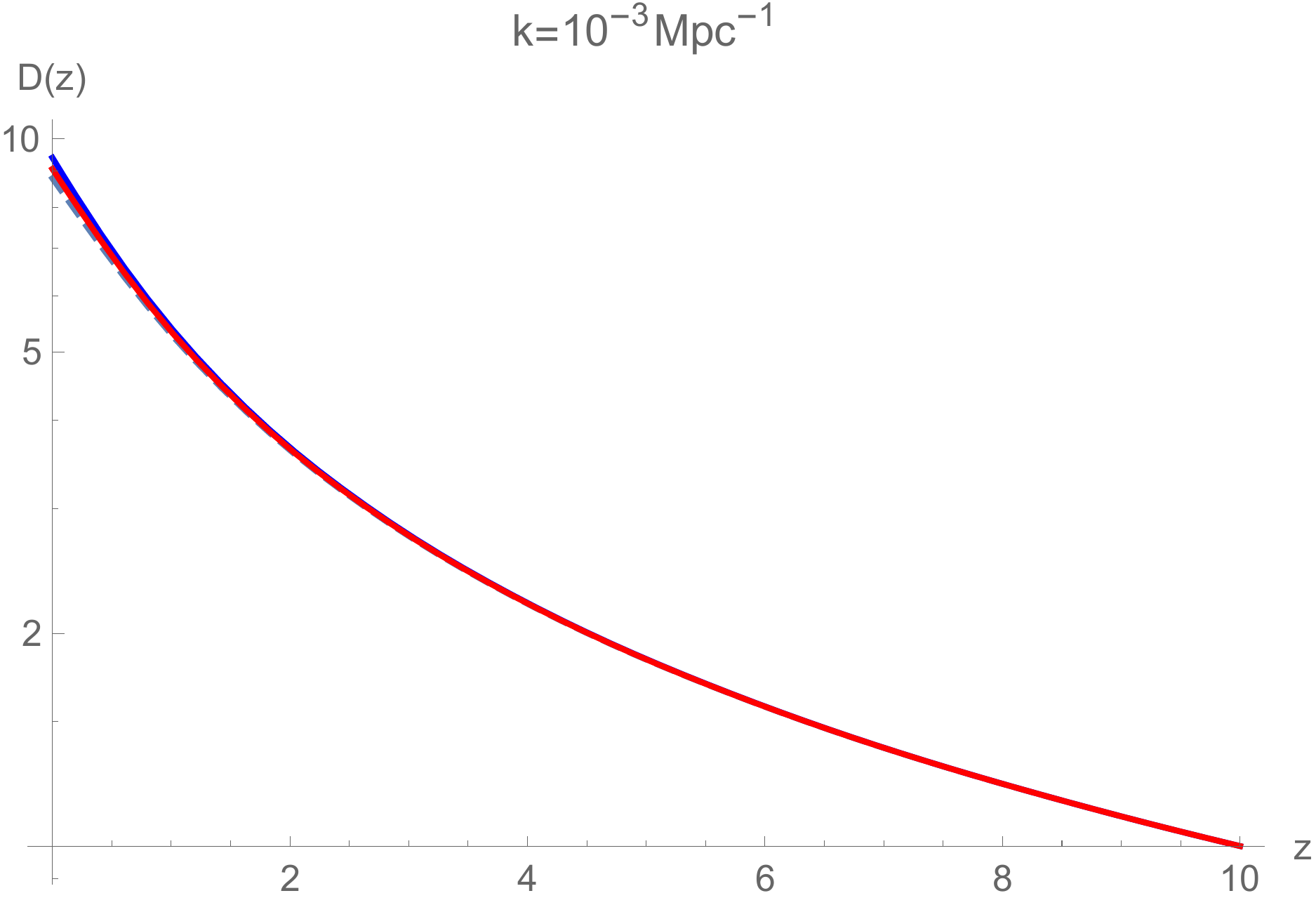}
\caption{  Growth functions.}
\label{Figgrth}
\end{minipage}
 \begin{minipage}{.5\textwidth} 
 \includegraphics[width=.8\linewidth]{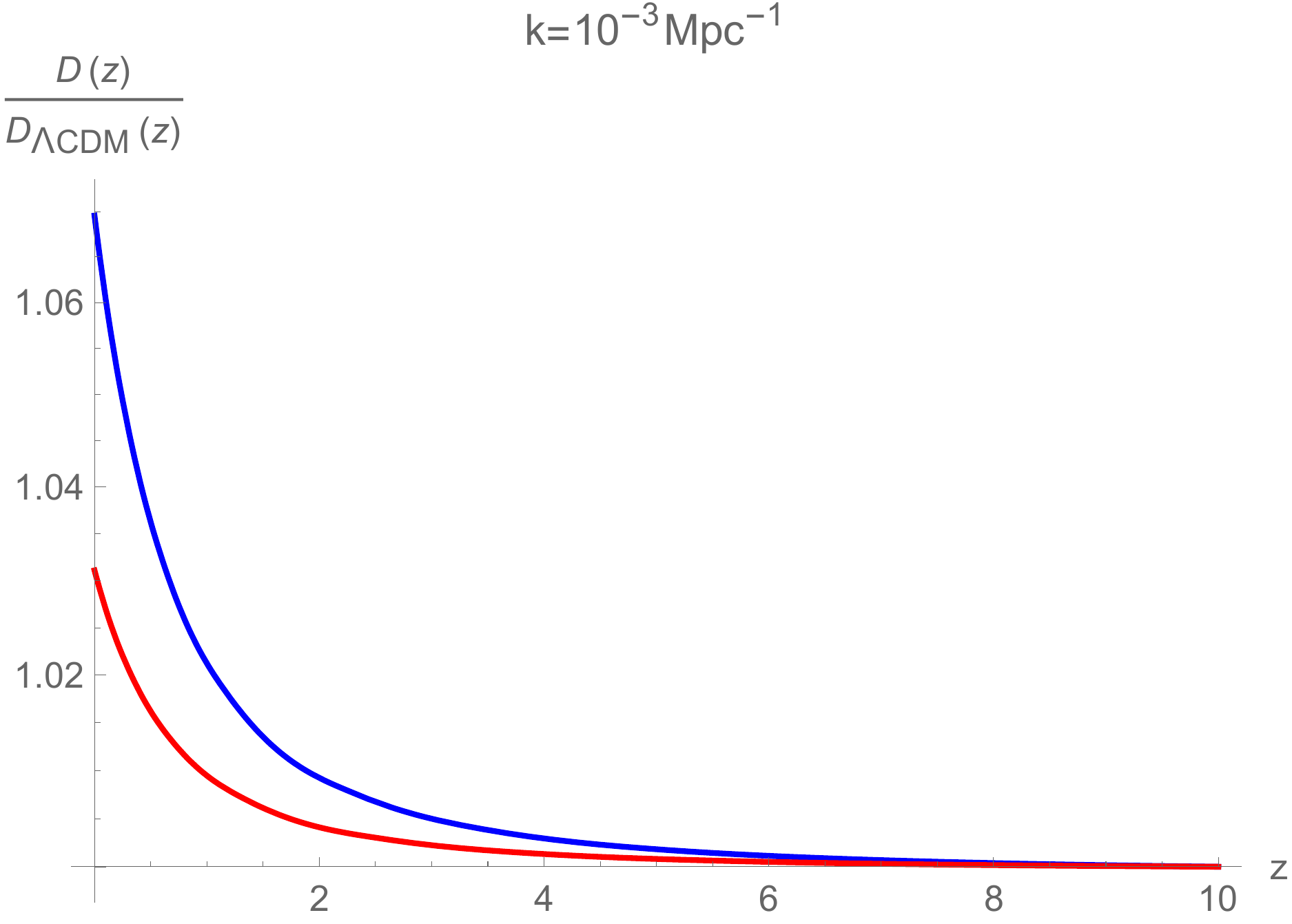}
 \caption{Ratio of the growth factor  with
   $\Lambda$CDM values; same colors as in the figure on the
   left.}
 \label{Figratio}
\end{minipage}
\end{figure}

 Clearly, in the absence of dark energy the growth function $D(z)$ will grow as $D(z)\sim 1+z$, while the presence of a
cosmological constant causes structure to grow less. This is not the
case for a $\Lambda$-medium where the growth of structure is
enhanced compared to standard  $\Lambda$CDM.
\\
Note that for $\Lambda$-media, differently from  $\Lambda$CDM, the contribution to the the growth function  from the entropy perturbations  is generically scale-dependent, and, since  $\delta\sigma(k)$ is constant in time, is sensitive to the $k$-dependence of the primordial perturbations specified by the entropic spectral index $n_\Lambda$.
\\
The  predictions for the  growth function  is shown in  Figure~\ref{Figgrth}, 
 where  the black
dotted curve shows the case of $\Lambda$CDM, while the blue dashed curve and the red curve
  show the growth function for  $\Lambda$-media with initial conditions  $n_\Lambda=1$ and $n_\Lambda=5$, respectively. The ratio of $D_{\delta_\Lambda}$ with the $\Lambda$CDM case is shown in Figure~\ref{Figratio}.

 We leave for a future work a complete
  numerical analysis suitable for  parameter estimation, but clearly
  the initial spectrum for $\delta \sigma$ is rather constrained from observations.

\section{Conclusions}
\label{CONC}

By using the EFT of self-gravitating
media, we have shown that there are stable media, protected by symmetries,
that  feature  an
exact (non-perturbative) equation of state of the form $p+\rho=0$, and
are physically   different from a CC.
Indeed, adiabatic
$\Lambda$-media and, in particular, 
$\Lambda$-solids and $\Lambda$-special super solids \footnote{Note
  that   the symmetry \eq{symm2} protects  the number of propagating DoF for the
class of Lagrangians $U(X,\;w_n)$   from quadratic higher derivative
operators~\cite{Dubovsky:2004sg,Rubakov:2008nh} and at the
non-perturbative level \cite{Comelli:2014xga}. }  exhibit  phonon-like fluctuations which, via
Einstein equations, induce non-trivial metric fluctuations. 
Moreover,
the Bardeen potentials and the density contrast of the sub-leading dark matter component $\delta_{m}$ grow as $a^2$,  in sharp contrast with the case of CC domination,  where the Bardeen potentials are decreasing and   $\delta_{m}$ is constant.
The presence of an intrinsic anisotropic stress induces a  strong
correlation  between the gravitational potentials: $\Psi=-2\;\Phi$ and
makes the spin two mode massive. Isentropic $\Lambda$-media are
characterised by frozen scalar perturbations, likewise a CC, though
spin two perturbations are non-trivial. 
Indeed, with the exception of $\Lambda$-perfect fluids, the dispersion
relation of  gravitational waves for stable $\Lambda$-media is the one
of a massive particle $\omega=\sqrt{k^2+m_g^2}$. 
These features can be  detectable in future dark energy surveys and
gravitational waves experiments. Already the linear matter power
spectrum gives  a tight constraint on the size of primordial
perturbations in the dark energy sector.
Some model in which nontrivial dark energy
  perturbations are present even when $w=-1$ was discussed in
 \cite{Amendola:2016saw} in light of the EUCLID mission; here we have
  given a model independent analysis of
  stability, the underlying symmetries and  the evolution of
  cosmological perturbations.
A detailed analysis of the phenomenological implications of
self-gravitating $\Lambda$-media with a parameters estimation will be carried out in
a separate work.

\section*{Acknowledgement}
We thank Sabino Matarrese for very useful discussions.
\appendix
\section{Mass parameters}
\label{mass-app}
The mass parameters $\{M_b\}$ are given by
\be
\begin{split}\label{MMI}
& M_0= \frac{\phi'{}^2}{2 \, a^4 \, \plm^2} \left[a^2 \left(U_{YY}-2
    \, U_X\right)-4\,  a \, \phi' \, U_{YX} +4\, \phi'{}^2 \, 
   U_{XX} \right] ;\\
&M_1 = \frac{2 \, \phi'{}^2}{\plm^2} \left[ a^{-2} \, U_X
  +\sum_{n=0}^3 a^{-4-2n} \, U_{y_n}\right] \, ;\\
&M_2 = -\frac{2}{\plm^2} \sum_{m=1}^3 n^2 \, U_{\tau_n} \, ;\\
&M_3 = \frac{1}{\plm^2} \left[
     2  \,\sum_{m,n=1}^3 m\, n\,
   a^{-2 \;(m+n)} \;U_{\tau _m \tau
   _n}
+2  \; \sum_{n=1}^3\left(n\;  a^{-3-2 \;n}\;
   U_{b \tau _n} 
-  \frac{ a^{-2 \;n}}{2}\; U_{\tau _n}\right)+
   \frac{1}{2\,a^6}\;U_{b^2}  \right] \, ;\\
&M_4=  \frac{\phi' }{\plm^2} \left[ \left( \sum_{m,n=1}^3 
   a^{1-2 n}\; U_{Y \tau
   _n}-\frac{U_Y}{2\;a}+  \frac{U_{bY}}{2\;a^4}\right) 
   -2\;
  \phi' \;\left(  \sum_{n=1}^3\,  a^{-2-2 n}
   \;U_{X \tau _n}+  \frac{U_{X}}{2\;a^2}- \frac{ U_{b X}}{2\;a^5}\,
  \right) \right]\, ;
\end{split}
\ee
In Minkowski space the mass parameters are obtained form the above
expression by setting $a=\phi'=1$.

\bibliographystyle{hunsrt}

\bibliography{fluidbiblio}

\end{document}